\documentclass[twocolumn,aps,pra,superscriptaddress,footinbib,showpacs]{revtex4-1}

\usepackage{graphicx,color} 
\usepackage{amsmath,amsthm}
\usepackage{amssymb}
\usepackage{bbm}
\usepackage{hyperref}
\newcommand{\bk}{\mathbf{k}}
\newcommand{\beps}{\boldsymbol{\epsilon}}
\newcommand{\bepss}{\boldsymbol{\epsilon}^*}
\newcommand{\bq}{\mathbf{q}}
\newcommand{\br}{\mathbf{r}}
\newcommand{\bR}{\mathbf{R}}

\newcommand{\sfT}{\hat{\mathsf{T}}}
\DeclareMathOperator{\tr}{tr}

\newcommand{\ls}{l} 
\newcommand{\Jg}{J_\text{g}}

\newcommand{\Je}{J_\text{e}}

\newcommand{\Meg}{M_\text{eg}} 
\newcommand{\sixj}[6]{\left\{\begin{array}{ccc}
                #1      & #2    & #3\\
                #4      & #5    & #6
                \end{array}\right\}}
\newcommand{\be}{\begin{equation}}
\newcommand{\ee}{\end{equation}} 
\newcommand{\avg}[1]{\overline{#1}}

\newcommand{\avgJ}[1]{\left\langle#1\right\rangle}
\newcommand{\textavgJ}[1]{\big\langle#1\big\rangle}
\newcommand{\bra}[1]{{\left\langle#1\right|}}
\newcommand{\ket}[1]{{\left|#1\right\rangle}}
\newcommand{\braket}[2]{{\left.{\left\langle#1\right|}#2\right\rangle}}

\newcommand{\INLN}{Institut Non Lin\'eaire de Nice, Universit\'e de Nice-Sophia Antipolis, CNRS, 06560 Valbonne, France}
\newcommand{\UKN}{Fachbereich Physik, Universit\"at Konstanz, 78457 Konstanz, Germany} 
\newcommand{\CQT}{Centre for Quantum Technologies, National University of
  Singapore, Singapore 117543, Singapore} 
\newcommand{\MM}{MajuLab, CNRS-UNS-NUS-NTU International Joint Research
Unit, UMI 3654, Singapore}
\newcommand{\NUS}{Department of Physics, National University of Singapore, Singapore 117542, Singapore}
\newcommand{\LKB}{Laboratoire Kastler Brossel, UPMC-Sorbonne Universit\'es, CNRS, ENS-PSL Research University, Coll\`{e}ge de France, 4 Place Jussieu, 75005 Paris, France}

\begin{document}
\author{Cord A.\ M\"uller}
\affiliation{\INLN} 
\affiliation{\UKN} 
\author{Beno\^it Gr\' emaud}
\affiliation{\MM} 
\affiliation{\CQT}
\affiliation{\NUS}
\affiliation{\LKB} 
\author{Christian Miniatura}
\affiliation{\MM} 
\affiliation{\CQT}
\affiliation{\NUS}
\affiliation{\INLN}
 
\date{\today}

\title{Speckle intensity correlations of photons scattered by cold atoms}

\begin{abstract}
The irradiation of a dilute cloud of cold atoms with a coherent light field produces a random intensity distribution known as laser speckle. Its statistical fluctuations contain information about the mesoscopic scattering processes at work inside the disordered medium. Following up on earlier work by Assaf and Akkermans [Phys.\ Rev.\ Lett.\  \textbf{98}, 083601 (2007)], we analyze how static speckle intensity correlations are affected by an internal Zeeman degeneracy of the scattering atoms. It is proven on general grounds that the speckle correlations cannot exceed the standard Rayleigh law. On the contrary, because which-path information is stored in the internal atomic states, the intensity correlations suffer from strong decoherence and become exponentially small in the diffusive regime applicable to an optically thick cloud. 
\end{abstract}


\maketitle

\section{Introduction}

In typical light-scattering experiments, a monochromatic light beam irradiates a medium with a certain wave vector and polarization (incident channel $a$), and the transmitted light intensity $T_{ab}$ is then measured for a certain final direction and polarization (scattered channel $b$). 
When the sample consists of randomly located scatterers, one observes a complex intensity pattern with large fluctuations known as a speckle pattern \cite{Goodman2007}. It originates from the interference between a large number of field amplitudes scattered by the sample. This pattern depends intricately on the positions of the scatterers, on the frequency and polarization of the incoming light, on the direction of observation, \textit{etc}.  

Since light scattering experiments allow for a precise control of the incident beam and accurate analysis of the scattered beam, they have been successfully used for thorough investigations of theoretical concepts originating in the field of mesoscopic electron transport, like weak and strong localization, and universal conductance fluctuations \cite{AkkermansMontambaux2007}. 
Already well documented for classical scatterers, light-scattering experiments investigating coherent multiple scattering have entered a new realm when cold atoms were used as scatterers \cite{Labeyrie1999}. It was rapidly realized that one crucial difference between classical and atomic scatterers resides in the atomic quantum internal structure, namely, the Zeeman degeneracy of the atomic dipole-transition levels \cite{Jonckheere2000,Mueller2001}. Each atom behaves like a freely orientable magnetic impurity that can exchange angular momentum with the scattered photon and thus store which-path information \cite{Miniatura2007}. 
Tracing out the atomic magnetic degrees of freedom then generally results in strong decoherence, as observed by a drastic reduction of the coherent backscattering signal \cite{Labeyrie2003} compared to the case of non-degenerate point scatterers \cite{Bidel2002}.  

To capture the generic features of a speckle pattern, one usually performs an average over spatial configurations of the scatterers. Typical quantities of interest are then the average transmission $\avg{T}_{ab}$ and the intensity-intensity correlation $\avg{T_{ab} T_{a'b'}}$ where the overbar denotes the spatial configuration average. 
Whenever the transmission $T_{ab}$ originates from the coherent superposition of a large number of uncorrelated, complex random amplitudes, the central limit theorem asserts that its probability distribution is the Rayleigh law, $P(T_{ab}) = \avg{T}_{ab}^{-1} \exp\{ -T_{ab}/\avg{T}_{ab}\}$  \cite{AkkermansMontambaux2007}. As as consequence, the statistical fluctuations obey $\avg{T^n_{ab}} = n! \avg{ T}_{ab} ^n$. 

Does the Rayleigh law hold also for the intensity-intensity correlations of light scattered by cold atoms? This question was first addressed in Ref.~\cite{Assaf2007}, where it was predicted that the Zeeman degeneracy leads to a drastic \emph{increase} of fluctuations in excess of the Rayleigh limit $\avg{T^2_{ab}} = 2 \avg{ T}_{ab} ^2$. This claim raised a controversy \cite{Gremaud2008,Assaf2008}. With this article, we present a detailed theoretical re-analysis of the protocol proposed in~\cite{Assaf2007}. In short, we ascertain that the internal Zeeman degeneracy reduces short-range speckle intensity correlations instead of enhancing them. 
 
The rest of the paper is structured as follows: In the next Sec.~\ref{sec:AmpIntCorr}, we state the relevant photon-scattering amplitude, determine the ensemble-averaged intensity, and derive the appropriate expression for the intensity correlations. 
In Sec.~\ref{sec:diff}, we compute the diffusive intensity correlations analytically as a function of Zeeman degeneracy, finding that they decay exponentially with the sample thickness for atoms with a degenerate internal ground state. Sec.~\ref{sec:Conc} concludes.

\section{Speckle correlations and the Rayleigh constraint}  
\label{sec:AmpIntCorr}

This section introduces the transition amplitude for the scattering of a photon from immobile atoms with internal structure, and determines the intensity and its correlations upon ensemble averaging over internal and external configurations. It is shown that the interference of amplitudes visiting atoms on different scattering paths can only involve Rayleigh transitions that preserve the internal state. Conversely, elastic Raman scattering events act as ideal which-path markers and thus reduce the relative weight of the interference terms.  Finally, we prove for the proposed setting that the relative fluctuations are rigorously bounded from above by the Rayleigh limit. 

\subsection{Light-scattering amplitudes}

Consider a dilute cloud of $N$ identical atoms, cooled to a
temperature low enough such that they can be considered immobile at
fixed classical positions $\bR = \{\br^j|j=1,\dots,N\}$, neglecting recoil effects and Doppler shifts.  The density is low enough for quantum denegeracy effects to be negligible.  
Then, each atom is an ideal classical point scatterer for electromagnetic
radiation. We look at light being scattered by a closed optical dipole transition between two internal electronic levels, with total angular momentum $\Jg$ 
in the ground state and $\Je$
in the excited state.      
In the absence of external fields, these levels are degenerate, and the ground state is spanned by magnetic quantum numbers $m$ in the range $-\Jg \leq m \leq \Jg$.   The set
of magnetic quantum numbers for all atoms is noted $M=
\{m^j|j=1,\dots,N\}$.  
We assume that the atomic sample is uncorrelated, and statistically homogeneous, in the following sense: 
\begin{enumerate}
\item[$(i)$] the spatial ensemble average 
$\avg{(\cdot)} = \int \prod_i \mathrm{d}\br^i f(\bR)$ 
over the classical positions is taken with the separable probability density 
\be
\label{avgR}
f(\bR) = \prod_i f(\br^i)   
\ee 
where generically $f(\br^i)=1/V$ with 
$V$ the sample volume accessible to each atom, and 
\item[$(ii)$] the ensemble average 
$\avgJ{\cdot} =  
\tr\{ \rho \cdot\}$ over the magnetic quantum numbers uses the direct product state 
\be
\rho =
 \bigotimes_{i=1}^N\rho^i,  
 \label{rhoproduct}
\ee
where 
$\rho^i =
\sum_{m^i} p_{m^i}\ket{m^i}\bra{m^i}  $
is the initial state of atom $i$.
\end{enumerate} 
In the present section, the initial populations $0\leq p_{m^i}\leq 1$ are left arbitrary. Only for the analytical calculations in the following Sec.~\ref{sec:diff}, all atoms are assumed to be in the completely isotropic state with $p_{m^i} = 1/(2\Jg+1)$.

Consider now the scattering of a monochromatic light beam by the atomic sample. 
An incident field mode with wave vector $\bk_a$ and 
transverse polarisation $\beps_a$ is scattered into a superposition of final modes 
$\bk_b,\beps_b$. The incident
light intensity is small compared to the saturation intensity of the
atomic transition, such that only the linear response, i.e.,
scattering of a single photon at a time, has to be considered. As a
consequence, the light is scattered elastically, $|\bk_a| = |\bk_b|
=k$.  
The internal state of an individual atom, however, couples to the
polarization of the scattered photon and thus can either remain the
same (Rayleigh transition) or change (degenerate Raman transition), with  
the concomitant change in the photon polarization imposed by angular-momentum 
conservation. 
The scattering amplitude from the initial state $\ket{\bk_a,\beps_a,M_a}$  to a final state 
$\ket{\bk_b,\beps_b,M_b}$ is the matrix element 
\be
A_{ab} = \bra{M_b} \sfT_{ab} \ket{M_a} 
\label{Aab}
\ee
of the transition operator (also called $T$-matrix \cite{Sheng2006}) $\sfT_{ab} = \bra{\bk_b,\beps_b}\sfT(\bR)\ket{\bk_a,\beps_a}$ acting on the internal states of all atoms \cite{Mueller2002}. 

Throughout the paper we assume that the atomic cloud satisfies the
diluteness conditions $n\lambda^3 \ll 1$ and $k\ls \gg 1$, where $n$
is the number density of atoms in the cloud, $\lambda=2\pi/k$ the
light wavelength, and $\ls$ the scattering mean free path. In this
case the light propagation can be described semiclassically in terms
of scattering paths $I=\{i_1,i_2,\dots\}$, 
defined as the ordered set of light-scattering atoms at positions $\{\br^i|i\in I\}=:\bR^I$. 
The probability amplitude \eqref{Aab}
then is the coherent superposition 
\be
A_{ab}=\sum_I A_{ab}^I. 
\label{AsumI}
\ee 
The diluteness condition also implies that one may safely discard recurrent scattering events, in 
which scatterers are visited repeatedly. Therefore, each atom
appears at most once in any given path $I$. 

Importantly, only an atom $i\in I$ 
in the actual scattering sequence can have its magnetic quantum number $m^i$ affected.  
All other atoms $j\in I^\mathrm{c}=\{j | j\notin I\}$ are bound to be mere spectators in this scattering sequence, and
their magnetic quantum numbers remain unchanged. In mathematical
terms, this is expressed by writing the transition operator for path $I$ 
in the form
$\sfT^I \otimes \mathbbm{1}^{I^\mathrm{c}}$, where $\sfT^I=\sfT(\bR^I)$ operates only on the scattering atoms, and $\mathbbm{1}^{I^\mathrm{c}}$ 
does nothing to the atoms in 
the spectator ensemble. 
By partitioning the magnetic quantum numbers $M = M^I \cup M^{I^\mathsf{c}}$ over  scatterers and spectators, we can write 
the scattering amplitude for the path $I$ as 
\be\label{AIab}
A_{ab}^I 
= \bra{M_b^I} \sfT^I_{ab} \ket{M_a^I}
\braket{ M_b^{I^\mathsf{c}} }{ M_a^{I^\mathsf{c}} } .   
\ee
The second factor imposes the constraint $m^j_b=m^j_a$ for all
spectator atoms $j\in I^\mathsf{c}$. An atom whose quantum number is not changed during the 
sequence $I$  can either be a member of $I$ and make a Rayleigh transition ($m_b^i=m_a^i$),
or it can be a member of the spectator set $I^\mathsf{c}$. But  
it is immediately clear that any atom that makes a Raman transition
($m^i_b \neq m^i_a$) must belong to a scattering sequence $I$. 
In other words, Raman transition events are path markers.

\subsection{Intensity}

The transition probability for the scattering process $a\to b$ is $|A_{ab}|^2 = A_{ab}A^*_{ab}$. 
Each of the amplitudes $A_{ab}$ here is the superposition  \eqref{AsumI} of path amplitudes,  
such that the total probability splits into a sum of
diagonal ($I=J$) and off-diagonal ($I\neq J$) contributions: 
\begin{eqnarray}
|A_{ab}|^2  &=& \sum_{I} A^I_{ab}A^{I*}_{ab} + \sum_{I\neq J} A^I_{ab}A^{J*}_{ab}. 
\label{A2IJ}
\end{eqnarray}
Because of the interference terms, this probability will in general depend quite
sensitively on the atomic positions, and on the light field vectors ($\bk_a,\beps_a,\bk_b,\beps_b$); the resulting random light intensity distribution is known as an optical speckle pattern. 
In the following, we first discuss the effect of the internal degrees of freedom, for a fixed spatial configuration, and then proceed to spatial averages. 

\subsubsection{Internal ensemble average}
\label{sec:internal}

The path amplitude \eqref{AIab} depends 
on the internal configurations, $M_a$ and $M_b$. 
We assume that these internal quantum numbers are not under microscopic 
control and trace them out by averaging over the initial $M_a$ with weights $p_{M_a}=\prod_i p_{m_a^i}$ and summing over the final $M_b$,  
\be
T_{ab} =  \sum_{M_a,M_b} p_{M_a}|A_{ab}|^2 = \avgJ{A_{ab}A^*_{ab}}.  
\label{TabAA}
\ee
Inserting the double path sum \eqref{A2IJ}  one thus arrives at the transition probability 
\begin{eqnarray}
T_{ab}  
 &=& \sum_I \avgJ{ A^{I}_{ab} A^{I*}_{ab}} + \sum_{I\neq J}  \avgJ{ A^{I}_{ab}A^{J*}_{ab}}. 
\label{TabIJ} 
\end{eqnarray}
Taking into account the product structure \eqref{AIab}, 
the internal averages are  
\begin{widetext}
\begin{eqnarray} 
\avgJ{A^{I}_{ab}A^{J*}_{ab}} 
&=& \sum_{M_a,M_b} p_{M_a} \bra{M_a^J}\sfT^{J\dagger}_{ab}\ket{M_b^J}\bra{M_b^I}\sfT^{I}_{ab}\ket{M_a^I} 
\braket{M_a^{J^\textsf{c}}}{M_b^{J^\textsf{c}}}\braket{M_b^{I^\textsf{c}}}{M_a^{I^\textsf{c}}}. 
\label{TJTIMM}
\end{eqnarray} 
\end{widetext}

There is an important difference between the diagonal, or incoherent contributions $(I=J)$ and the coherent contributions $(I\neq J)$ to the total intensity \eqref{TabIJ}. This difference is best appreciated in the single-scattering approximation (valid for an optically thin medium, such as a collection of a few scatterers), where the scattering path $I\mapsto i$ reduces to the visited atom
\cite{LuYinPhD2012}. In this situation, using \eqref{TJTIMM} with $I=J=i$, one sees that the incoherent contribution of atom $i$, 
\begin{eqnarray} 
\avgJ{A^{i}_{ab} A^{i*}_{ab}} &=& 
\sum_{m_a^i,m_b^i} p_{m_a^i} \bra{m_a^i}\sfT^{i\dagger}_{ab} \ket{m^i_b}\bra{m^i_b} \sfT^{i}_{ab}\ket{m_a^i} \quad 
\label{TiTi}
\end{eqnarray} 
involves all transitions between its internal states that are allowed by the dipole selection rules. Notably, these can include degenerate Raman transitions ($m^i_b\neq m^i_a$). The internal quantum numbers of the $N-1$ spectator atoms are irrelevant and disappear by the elementary normalization $\sum_{m_a^j,m_b^j} p_{m_a^j} |\langle m_a^j|m_b^j\rangle|^2 = \tr\{\rho^j\} = 1$. 
In contrast, in the coherent contribution from two different atoms,    
\begin{eqnarray} 
\avgJ{A^{i}_{ab} A^{j*}_{ab}}  &=& \!\!\! \sum_{m_a^i,m_a^j}\!\!\! p_{m_a^i}p_{m_b^i} \bra{m_a^j}\sfT^{j\dagger}_{ab} \ket{m^j_a} \bra{m^i_a} \sfT^{i}_{ab}\ket{m_a^i}, \qquad 
\label{TiTj}
\end{eqnarray}
each of the two atoms  $i\neq j$ belongs to the spectator set of the other ($i\in J^{\textsf{c}}$ and $j\in I^{\textsf{c}}$). Consequently, two overlap factors $\braket{m_a^i}{m_b^i}\langle m_b^j|m_a^j\rangle$ enforce Rayleigh transitions for both atoms $i$ and $j$ in the step from Eq.~\eqref{TJTIMM} to Eq.~\eqref{TiTj}, and the $N-2$ remaining atoms in their joint spectator set $(I\cup J)^\textsc{c}$ are traced out as irrelevant. 

The preceding reasoning, together with \eqref{TiTi} and \eqref{TiTj} written again with $i\mapsto I$ and $j\mapsto J$, holds also for multiple scattering  sequences, as long as the two paths $I,J$ do not share a common scatterer, $I\cap J=\emptyset$,  such that $I\subset J^{\textsf{c}}$ and $J\subset I^{\textsf{c}}$. 
This is actually the dominant situation in a dilute cloud with $N\gg1$ atoms. 
For future use, we note that the coherent internal ensemble average then decouples as 
\be\label{avTiTjfactors}
\avgJ{A^{I}_{ab} A^{J*}_{ab}} =  \avgJ{\sfT^{I}_{ab}} \avgJ{\sfT^{J}_{ab}}^* , \qquad (I\cap J= \emptyset), 
\ee
where $\textavgJ{\sfT^I_{ab}} = \tr\{\rho \sfT^I_{ab} \}$ involves only atoms of path $I$.

It is a crucial insight for this paper that the internal path markers impose the Rayleigh-scattering constraint for the interference terms, whereas the incoherent terms now include also the Raman scattering. 
This obviously complies with the venerable rule of scattering theory (inherited from the superposition principle of quantum mechanics) that amplitudes for indistinguishable processes should be added coherently, whereas the probabilities for distinguishable processes must be added incoherently. And indeed, as already indicated above, as soon as a single atom makes a Raman transition, this event acts as a path marker that renders the amplitude distinguishable from all the others where this atom was not visited. 
Thus, the incoherent processes, including the Raman scattering events, are much more numerous compared to the non-degenerate case. As a consequence, the Rayleigh-scattering constraint reduces the apparent speckle contrast with respect to this enhanced background.  

Of course, the reduction of interference visibility in the presence of which-path information 
is not a new insight special to mesoscopic systems, but has on the contrary been very well studied in numerous contexts. Prominent examples include the quantum-optical realization of Young's double-slit experiment using two trapped ions with internal structure \cite{Eichmann1993,Itano1998}, Mach-Zehnder photon interferometry with path encoding by polarization \cite{Schwindt1999}, and matter-wave interferometry with path encoding in internal electronic states \cite{Durr1998,Durr1998a}.

\begin{figure*}
\begin{center}
\includegraphics[width=0.3\textwidth]{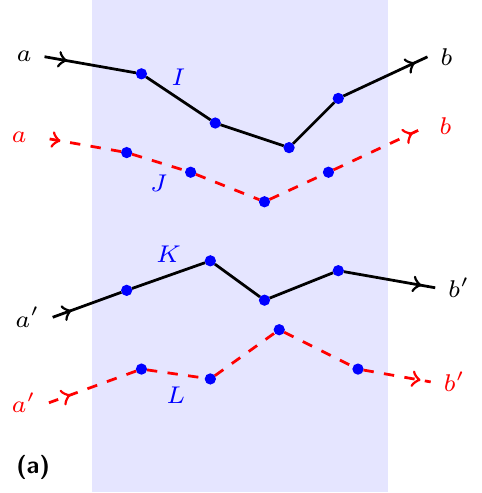}\hfill
\includegraphics[width=0.3\textwidth]{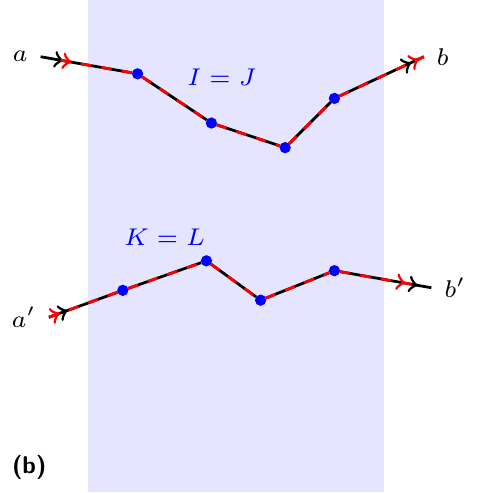}\hfill
\includegraphics[width=0.3\textwidth]{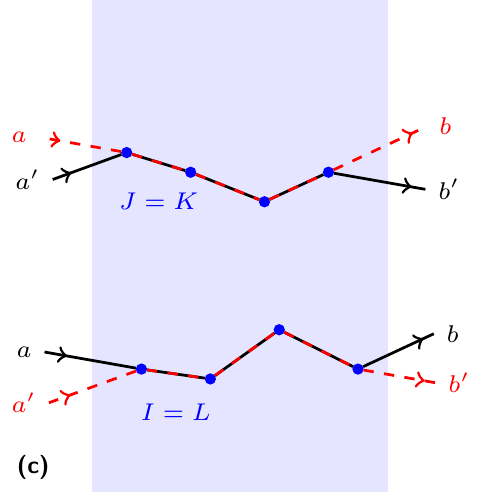}
\caption{(Color online) Amplitudes for light scattering from mode $a=(\bk_a,\beps_a)$ to $b=(\bk_b,\beps_b)$ in a first pulse and from  $a'$ to $b'$ in a second pulse, across a dilute cloud of immobile point scatterers. 
 \textsf{\textbf{(a)}} Possible paths $I$, $J$, $K$, $L$, drawn with 4 representative scatterers each.   
\textsf{\textbf{(b)}} Pairing $I=J$ and $J=K$, yielding the product $\avg{T}_{a b}\avg{T}_{a'b'}$ of average intensities.    
\textsf{\textbf{(c)}} Pairing $I=L$ and $J=K$, describing intensity correlations $\avg{\delta T_{ab} \delta T_{a'b'}}$ .}
\label{fig:paths} 
\end{center} 
\end{figure*}

\subsubsection{Spatial ensemble average}

Consider again the path-resolved transition probability \eqref{TabIJ} for light scattering in transmission across a dilute cloud of scatterers, as depicted in the upper half of Fig.~\ref{fig:paths}(a). In the interference terms, all path pairs $I\neq J$ involve large, random phase differences. (In a transmission geometry, one can neglect the pairing of time-reversed paths that is responsible for coherent backscatting. As pointed out before, one can also discard subdominant contributions where $I$ and $J$ share common scatterers.) Under the spatial ensemble average over random atomic positions, the interference terms $I\neq J$ vanish, and thus the average intensity contains only the paired paths $I=J$ of the incoherent contribution, 
\be
\avg{T}_{ab} = \avg{ 
  \sum_I \avgJ{\sfT^{I\dagger}_{ab} \sfT^{I}_{ab}}, 
}
\label{avgTab} 
\ee
as depicted in the upper half of Fig.~\ref{fig:paths}(b). 
This average transmitted intensity in the far field is a rather structureless, smooth function of wave vectors and polarization vectors without mesoscopic signatures of coherence. More interesting information can be gained from intensity correlations, to be discussed next.

\subsection{Intensity correlations} 

We now consider the protocol proposed in \cite{Assaf2007} and analyze the correlations between the two speckle patterns obtained in consecutive scattering sequences $a\to b$  and $a' \to b'$, recorded one shortly after the other. 
The waiting time between pulses is supposed to be short enough for the
atomic positions to remain unchanged. Then, after each double pulse, the atoms slowly move to new positions.  This two-pulse experiment is performed many times, until the ensemble average $\avg{(\cdot)}$ over random atomic
positions is achieved.


Starting from the connected intensity correlator  
\be \label{defCab}
\avg{\delta T_{ab} \delta T_{a'b'}}= \avg{T_{ab} T_{a'b'}} -  \avg{T}_{ab} \avg{T}_{a'b'},
\ee
one defines the magnitude of fluctuations compared to the background,  
\be
C_{ab,a'b'} = \frac{\avg{\delta T_{ab} \delta T_{a'b'}}}{ \avg{T}_{ab} \avg{T}_{a'b'}}, 
\label{Cababdef}
\ee
as the quantity of central interest. The product of intensities \eqref{TabAA} before spatial averaging reads 
\be
T_{ab}T_{a'b'} = \sum_{IJKL} \avgJ{A_{ab}^I A^{J*}_{ab}} \avgJ{ A^{K}_{a'b'}A_{a'b'}^{L*}}. 
\ee
Along with Ref.~\cite{Assaf2007}, we assume here that the atoms are in the same, completely uncorrelated state \eqref{rhoproduct} at the arrival of each pulse, such that the averages over the internal configurations decouple. 
Fig.~\ref{fig:paths}(a) represents these amplitudes schematically. 
Just as for the average intensity \eqref{avgTab}, the only terms surviving the spatial average are those where  the phases accumulated between different atoms are prefectly compensated.
This leaves two possible path pairings, namely $(I=J,K=L)$
and $(I=L,J=K)$, depicted in Fig.~\ref{fig:paths}(b) and (c), respectively. 

Following \cite{Assaf2007}, we assume that the paths $I$ and $K$ do not share any scatterer ($I\cap K=\emptyset$).%
\footnote{Long-range correlations originating from shared scatterers do exist, but they are of higher order in $1/k\ls$ \cite{AkkermansMontambaux2007} and will be disregarded throughout this work.}  
The first pairing  $(I=J,K=L)$ then yields  
the product of the average intensities, 
\be
\avg{ \sum_{I} \avgJ{A_{ab}^I A^{I*}_{ab}}}\  \avg{ \sum_{K} \avgJ{A^{K}_{a'b'}A_{a'b'}^{K*}}}
= \avg{T}_{ab}\avg{T}_{a'b'}. 
\ee 
As a consequence, the speckle correlations \eqref{defCab} are entirely due to the second pairing $(I=L,J=K)$:
\begin{eqnarray}
\label{delTdelT}
\avg{\delta T_{ab} \delta T_{a'b'}}  = \avg{ 
\sum_{IK}
 \avgJ{A_{ab}^{I} A_{ab}^{K*}} \avgJ {A_{a'b'}^{K} A_{a'b'}^{I*}}}. 
\end{eqnarray}
Since $I\cap K=\emptyset$, both external and internal averages decouple. In particular, Eq.~\eqref{avTiTjfactors} can be used, resulting in  
\begin{eqnarray} 
\avg{\delta T_{ab}\delta T_{a'b'}}
&=& \avg{ \sum_{I} \avgJ{\sfT_{ab}^I} \avgJ{\sfT_{a'b'}^I}^*} \, \avg{\sum_K \avgJ{\sfT_{a'b'}^K} \avgJ{\sfT_{ab}^K}^*} \qquad \\
&=&  \left| \avg{ \sum_{I} \avgJ{\sfT_{ab}^I} \avgJ{\sfT_{a'b'}^I}^*}    
\right|^2.
 \label{delTdelTfactored}
\end{eqnarray}
The derivation of this expression  is the key achievement of our paper. In essence, we find that under the assumption of uncorrelated distributions, the mesoscopic speckle correlations are given by a path-averaged propagator involving products of independent internal averages $ \textavgJ{\sfT_{ab}^I} \textavgJ{\sfT_{a'b'}^I}^*$. As explained in Sec.~\ref{sec:internal} above, this is ultimately a consequence of the Rayleigh scattering constraint for interference contributions. Ref.~\cite{Assaf2007} predicted unphysically large speckle correlations  essentially because the Rayleigh constraint was not taken into account; herein lies the difference between Eq.~(5) in the Reply by Assaf and Akkermans \cite{Assaf2008} and our result \eqref{delTdelTfactored}.

\subsection{The Rayleigh bound}

We proceed to prove on general grounds that under the stated hypotheses (disjoint paths $I\cap K=\emptyset$ and uncorrelated, but otherwise arbitrary distributions of internal and external degrees of freedom), the relative fluctuations \eqref{Cababdef} are bounded from above by the Rayleigh law, 
\be\label{Cbound}
C_{ab,a'b'} \leq 1.   
\ee
\textit{Proof:} In a first step, let  
\be
 \avg{\sum_I\avgJ{\sfT_{ab}^I}^*\avgJ{\sfT_{a'b'}^I}} =:\left(\sfT,\sfT'\right), 
\ee
denote a scalar product with respect to the trace over the atomic external degrees of freedom. The speckle correlations \eqref{delTdelTfactored} are then bounded by the Cauchy-Schwartz inequality  
\be
\avg{\delta T_{ab}\delta T_{a'b'}} = 
\left|\left(\sfT,\sfT'\right)\right|^2 \leq \left(\sfT,\sfT\right) \left(\sfT',\sfT'\right). 
\ee
Equivalently,
the general correlator \eqref{Cababdef} cannot be larger than the geometric average of the diagonal correlators  $C_{ab,ab}=|\big(\sfT,\sfT\big)/\avg{T}_{ab}|^2$: 
\be
C_{ab,a'b'} \leq  \sqrt{C_{ab,ab}C_{a'b',a'b'}}. 
\label{Cabboundgeomavg}
\ee
In a second step, we prove that each diagonal correlator obeys the Rayleigh bound,    
\be
C_{ab,ab} = 
 \left[
\frac{
  \avg{ \sum_I \avgJ{\sfT_{ab}^I}^* \avgJ{\sfT_{ab}^I}}
  } 
  {
    \avg{ \sum_I \textavgJ{ \sfT_{ab}^{I\dagger} \sfT_{ab}^I}}
}
\right]^2 \leq 1. 
\label{Cabbound}
\ee  
This bound is a corollary of the Cauchy-Schwarz inequality 
\be
|{\avgJ{A^\dagger B}}|^2 \leq \avgJ{A^\dagger A}\avgJ{B^\dagger B},
\ee 
now for the scalar product $\avgJ{A^\dagger B}= \tr\{ \rho A^\dagger B\}$ of the trace over the internal degrees of freedom. 
Setting $A=\mathbbm{1}$ and $B=\sfT^I_{ab}$, one has
\be
\label{CSineq}
\left |{\avgJ{\sfT_{ab}^I}}\right|^2  \leq \left\langle \sfT_{ab}^{I\dagger} \sfT_{ab}^I \right\rangle , 
\ee 
and by linearity of the path sum and ensemble average,
\be
\avg{\sum_I \left|{\avgJ{\sfT_{ab}^I}}\right|^2} \leq 
 \avg{\sum_I \left\langle \sfT_{ab}^{I\dagger} \sfT_{ab}^I\right\rangle},
\label{CSpaths}
\ee
whence \eqref{Cabbound}. Together with \eqref{Cabboundgeomavg}, this yields the Rayleigh bound \eqref{Cbound}. \hfill $\blacksquare$

Clearly, expression \eqref{Cabbound} shows that a necessary and sufficient condition for satisfying the Rayleigh law $C_{ab,ab}=1$ is that the internal ensemble average factorizes, 
$\textavgJ{ \sfT_{ab}^{I\dagger} \sfT_{ab}^I }=  \big|\textavgJ{\sfT_{ab}^I }\big|^2 $. 
And we know from Ref.~\cite{Mueller2001}--- at least for the isotropic internal state considered  in \cite{Assaf2007}---that the average intensity is equal to the averaged amplitudes squared if and only if the ground state is non-degenerate ($\Jg=0$). Therefore, we can formally conclude that the fluctuations reach their maximum value, $C_{ab,ab}=1$, only for $\Jg=0$. In all other cases $\Jg>0$, the fluctuations are reduced, $C_{ab,ab}<1$.  

\section{Diffusive intensity and speckle correlations}
\label{sec:diff} 

In the following, we re-calculate the speckle correlations along the lines of
\cite{Assaf2007}, but with the ensemble average over internal quantum numbers as defined in Eq.~\eqref{delTdelTfactored}. 
We first recall how to compute the average light intensity scattered by a single atom with internal degeneracy
 in the isotropic ground state 
\be
\rho_0^i = \frac{1}{2\Jg+1}\sum_{m^i}\ket{m^i}\bra{m^i}.
\label{rho0i}
\ee
Then 
we determine the single-scattering vertex for the correlation propagator. In a second step, we compute the diffusive intensities and their correlations. Finally, we determine the relative fluctuations observed in transmission across an optically thick  slab. 

\subsection{Single-scattering vertices} 

The dipole scattering of a photon from mode $(\bk_a,\beps_a)$ to mode $(\bk_b,\beps_b)$ by a single
atom at position $\br^i$ is described by the scattering operator 
\be
\sfT^i_{ab} = (\bepss_b \cdot \hat{\mathbf{d}})(\hat{\mathbf{d}}\cdot \beps_a) e^{i(\bk_a-\bk_b)\cdot\br^i}. 
\label{T1ab} 
\ee
Proportionality factors that are irrelevant in the present context have been set to unity, and 
$\hat{\mathbf{d}}$ is the reduced atomic dipole operator connecting the ground-state manifold with angular momentum $\Jg$ to the excited-state manifold with angular momentum $\Je$. For the isotropic internal state \eqref{rho0i}, the average single-scattering intensity  $\tr\{\rho_0^i \sfT^{i\dagger}_{ab} \sfT^i_{ab}\} = \textavgJ{\sfT_{ab}^{i\dagger} \sfT_{ab}^i}$  
can be evaluated in closed form using the techniques of irreducible 
tensor operators \cite[Eq.~(39)]{Mueller2001}: 
\be\label{T1intensity}
\avgJ{ \sfT^{i\dagger}_{ab} \sfT^i_{ab}}
=\Meg\left[ w_1 |\beps_a\cdot\bepss_b|^2 +
w_2 |\beps_a\cdot\beps_b|^2 + w_3\right].
\ee
Here, 
\be
\Meg =  \frac{2\Je+1}{3(2\Jg+1)}
\label{Megdef}
\ee 
denotes a ratio of level degeneracies, and the weights $w_j$ are rotational invariants that depend only on the total angular momenta $\Jg$, $\Je$. 
It is advantageous to define the reduced vertex function $I(\beps_a,\bepss_b,\beps_b,\bepss_a) = \Meg^{-1}  \textavgJ{\sfT_{ab}^{1\dagger} \sfT_{ab}^1}$ given by 
\be\label{Ivertex}
I(\beps_a,\bepss_b,\beps_b,\bepss_a) = w_1 |\beps_a\cdot\bepss_b|^2 +
w_2 |\beps_a\cdot\beps_b|^2 + w_3. 
\ee
This function is essentially the radiation pattern, proportional to the
differential scattering cross section, in the given polarization channel $\beps_a \to \beps_b$.  


Let us now examine the single-scattering building block for the correlation propagator  \eqref{delTdelTfactored}, namely 
\be
I^{(\text{c})}_{ab,a'b'} =  \Meg^{-1} \avgJ{\sfT_{ab}^i} \avgJ{\sfT_{a'b'}^i}^*
\label{Icab}
\ee  
In the bulk of the scattering medium, one has $\bk_a=\bk_{a'}$ and $\bk_b=\bk_{b'}$ [cf.\ Fig.~\ref{fig:paths}(c); the dependence on the external momenta will be re-established below], and we can focus on the dependence on polarization vectors. Because of the Rayleigh-scattering constraint, the correlation vertex \eqref{Icab}
is proportional to the product of two independent traces over internal states.
Each trace selects the scalar component $\kappa=0$ of the dyadic tensor operator $\hat{\mathbf{d}}\hat{\mathbf{d}}$---a considerable simplification compared to the single-scattering intensity vertex \eqref{Ivertex} which couples components $\kappa=0,1,2$---and results in 
\be\label{Icorr}
I^{(\text{c})}(\beps_a,\bepss_b,\beps_{b'},\bepss_{a'}) = \Meg
(\beps_a \cdot \bepss_b)(\beps_{b'}\cdot\bepss_{a'}).
\ee 
The scalar products between polarization vectors here mean that the atom radiates exactly the polarization it has absorbed, in both sequences $a\mapsto b$ and $a'\mapsto b'$, 
as required by angular-momentum conservation under the Rayleigh-scattering constraint. In
this sense, the speckle correlations described by Eq.~\eqref{delTdelTfactored} originate from a much more restricted set
of events than the full intensity. We are thus entitled to anticipate
that internal degeneracies will rather reduce the observable speckle
correlations, instead of enhancing them.   

\subsection{Multiple scattering: the diffuson} 

In order to sum the multiple-scattering intensity series, it has proven useful  \cite{Mueller2002}
to write the cartesian elements 
$I_{il;jk} = w_1 \delta_{ij}\delta_{kl} + w_2 \delta_{ik}\delta_{jl} +
w_3 \delta_{il}\delta_{jk}$ 
of the single-scattering 4-point vertex \eqref{Ivertex} as a sum of its irreducible tensorial
components in the incident and scattered polarization channels, respectively: 
\be\label{Idecomp}
I_{il;jk} = \sum_{\kappa=0}^{2} \lambda_\kappa T^{(\kappa)}_{il;jk}.
\ee 
The projectors onto the three subspaces 
$\kappa=0,1,2$ of
dimension $2\kappa+1$ describing population, orientation, and alignment, respectively   
\cite{Omont1977}, are in Cartesian elements  
\begin{eqnarray}
T^{(0)}_{il;jk} &=& \frac{1}{3} \delta_{il}\delta_{jk},\label{TK0} \\
T^{(1)}_{il;jk} &=& \frac{1}{2}(\delta_{ij}\delta_{kl} - \delta_{ik}\delta_{jl}), \label{TK1}\\
T^{(2)}_{il;jk} &=& \frac{1}{2}(\delta_{ij}\delta_{kl} + \delta_{ik}\delta_{jl}) -  \frac{1}{3}
\delta_{il}\delta_{jk}, \label{TK2}
\end{eqnarray} 
and the corresponding 
eigenvalues can be expressed by $6j$-symbols of 
angular momentum theory: 
\be \label{lambdaK}
\lambda_\kappa   = 3(2\Je+1) \sixj{1}{1}{\kappa}{\Je}{\Je}{\Jg}^2. 
\ee

Using the decomposition \eqref{Idecomp}, the entire multiple scattering series
for the average intensity has been summed exactly in Ref.~\cite{Mueller2002}, including the 
constraint of transversality for the far-field photons
exchanged between distant atoms. 
In the diffusion approximation, the bulk intensity
propagator (known as the ``diffuson'') for the polarization channel $\beps_a\mapsto \beps_b$ 
reads in Fourier representation 
\be\label{Gammadiff}
\Gamma_{ab}(\bq) = \sum_{\kappa=0}^2 \frac{t^{(\kappa)}_{ab}}{q^2 + (D\tau_\kappa)^{-1}}. 
\ee
Here, the momentum $\bq$ is the Fourier variable conjugate to the position difference, and 
the mode populations 
\be 
t^{(\kappa)}_{ab} =
\sum_{ijkl} \beps_{ai}\bepss_{bj}\beps_{bk}\bepss_{al}T^{(\kappa)}_{il;jk}
\label{tKab}
\ee
depend only
on the initial and final polarization. Each mode has a
diffusive propagator with a characteristic
extinction range $\ls_\kappa^2 = D\tau_\kappa$. Here, $D =
\ls^2/3\tau$
is the (scalar) diffusion constant with $\ls$ the mean free path and $\tau$
the energy transport time \cite{Mueller2002}. 
The depolarization times      
\be\label{tauK}
\tau_\kappa = \frac{\lambda_\kappa \tau }{3 (1 - b_\kappa \lambda_\kappa)}
\ee
describe the damping of mode populations. This depolarization is due to two effects: First, the transversality constraint of free-space propagation between
atoms, as encoded by the photon propagator eigenvalues 
\be 
b_0=1,\quad b_1=\frac{1}{2},\quad  b_2=\frac{7}{10}. 
\ee
Second, a scrambling of polarization by internal atomic Raman transitions, as encoded in the vertex eigenvalues $\lambda_\kappa$, Eq.~\eqref{lambdaK}. 

Only the scalar mode $\kappa=0$ that counts the
total intensity or number of photons has an infinite
lifetime, $\tau_0^{-1}=0$, since $b_0\lambda_0=1$ for all $\Je,\Jg$. 
The two other modes $\kappa=1,2$ have $b_\kappa\lambda_\kappa<1$ for all $\Jg,\Je$
and thus describe depolarization on the scale of the transport 
time $\tau$.    

\subsection{Multiple scattering: the correlon}

The previous derivation for the diffuson can be applied to the correlation
propagator of \eqref{delTdelTfactored} (dubbed henceforth ``correlon''), by 
substituting the correlation vertex \eqref{Icorr} for the 
intensity vertex \eqref{Ivertex}.  The correlation vertex
decomposes as 
\be
I_{il;jk}^{(\text{c})} = \sum_{\kappa=0}^{2} \lambda_\kappa^{(\text{c})}
T^{(\kappa)}_{il;jk} 
\ee 
over the same projectors \eqref{TK0}--\eqref{TK2}. Bot now there is only a single, fully degenerate eigenvalue for all modes $\kappa=0,1,2$: 
\be
\lambda_\kappa ^{(\text{c})}= \Meg = \frac{2\Je+1}{3(2\Jg+1)}. 
\ee 
This is nothing but 
a complicated way of saying that the vertex \eqref{Icorr} is proportional to the identity, preserving the polarization under Rayleigh scattering.  
The diffusive correlation propagator then takes 
the same form as \eqref{Gammadiff}, 
\be
\Gamma_{ab,a'b'}^{(\text{c})}(\bq) =
\delta_{\bk_a-\bk_b,\bk_{a'}-\bk_{b'}} \sum_{\kappa=0}^2
\frac{t^{(\kappa)}_{ab,a'b'}}{q^2 + (D\tau_\kappa^{(\text{c})})^{-1}}. 
\ee
The prefactor expresses the conservation of total momentum, restored
by the ensemble average over atomic positions, 
for the four independent momenta entering the correlator. 
The mode occupation
factors also depend on all four polarizations concerned, 
\be
t^{(\kappa)}_{ab,a'b'} = \sum_{ijkl} 
\beps_{ai}\bepss_{bj}\beps_{b'k}\bepss_{a'l}T^{(\kappa)}_{il;jk}. 
\ee
In the parallel channels $\beps_{a'}=\beps_a$ and $\beps_{b'}=\beps_b$, one finds $t^{(\kappa)}_{ab,ab}=t^{(\kappa)}_{ab}$ as given by \eqref{tKab}. 
The decorrelation times 
\be\label{tauKcorr} 
\tau^{(\text{c})}_\kappa = \frac{\Meg \tau }{3 (1 -
  b_\kappa \Meg)}
\ee
are non-negative under all circumstances because $0\leq b_\kappa \leq 1$ and $0\le \Meg \le 1$ for all $\Je,\Jg$ because of the dipole selection rule $|\Je-\Jg|\le 1$. 

Only atoms with a non-degenerate ground level ($\Jg=0,\Je=1$) act as scalar point scatterers, and
there is no difference between the average intensity and the average
amplitude squared (as discussed at length in the context of coherent
backscattering \cite{Mueller2001}), such that  $ \textavgJ{\sfT_{ab}^\dagger \sfT_{ab}} =  \textavgJ{\sfT_{ab}}^* \textavgJ{\sfT_{ab}}$. Equivalently, the correlation
vertex and the intensity vertex have the same eigenvalues,
$\lambda_\kappa=\lambda_\kappa^{(\text{c})}=1$. Therefore, from Eq.~\eqref{Cabbound} we recover the Rayleigh law of unit relative fluctuations, 
\be\label{Rayleigh} 
C_{ab,ab} =  1 \qquad (\Jg=0).  
\ee

But in general, contrary to
what is claimed in \cite{Assaf2007}, we find no negative
decorrelation times that could signal an instability and a build-up of
correlations over large distances.  For degenerate dipole scatterers with $\Jg>0$, the correlation vertex
eigenvalues differ from their intensity counterparts, but, as shown by \eqref{CSineq}, the speckle correlations can only be reduced by the internal degeneracy, never enhanced. 

\subsection{Speckle correlations across a slab} 

As an application of the preceding formalism, we compute the
forward correlations $a=a'$, $b=b'$ 
measured in transmission across a slab of width $x=L/\ls$
in units of the mean free path $\ls$, following \cite{Assaf2007}. 
The solution of the diffusion equation in the slab, for each irreducible polarization mode $\kappa$, is   
\be\label{DKofx}
\Gamma_\kappa(x) = \frac{(\sinh\gamma_\kappa)^2}{\gamma_\kappa\sinh \gamma_\kappa x} 
\ee
with reduced relaxation rates 
\be \gamma_\kappa = \sqrt{\tau/\tau_\kappa} = \sqrt{3(\lambda_\kappa^{-1} - b_\kappa)}
\label{gamma0c}
\ee 
for
both the intensity (whose eigenvalues $\lambda_\kappa$ are given by \eqref{lambdaK}), and
the correlation (whose eigenvalues are $\lambda_\kappa^{(\text{c})}=\Meg$). 
For conserved modes with $\gamma_\kappa=0$, \eqref{DKofx} reduces to
$\Gamma_\kappa(x) = 1/x$. 

\begin{figure} 
\begin{center}
\includegraphics[width=0.85\linewidth]{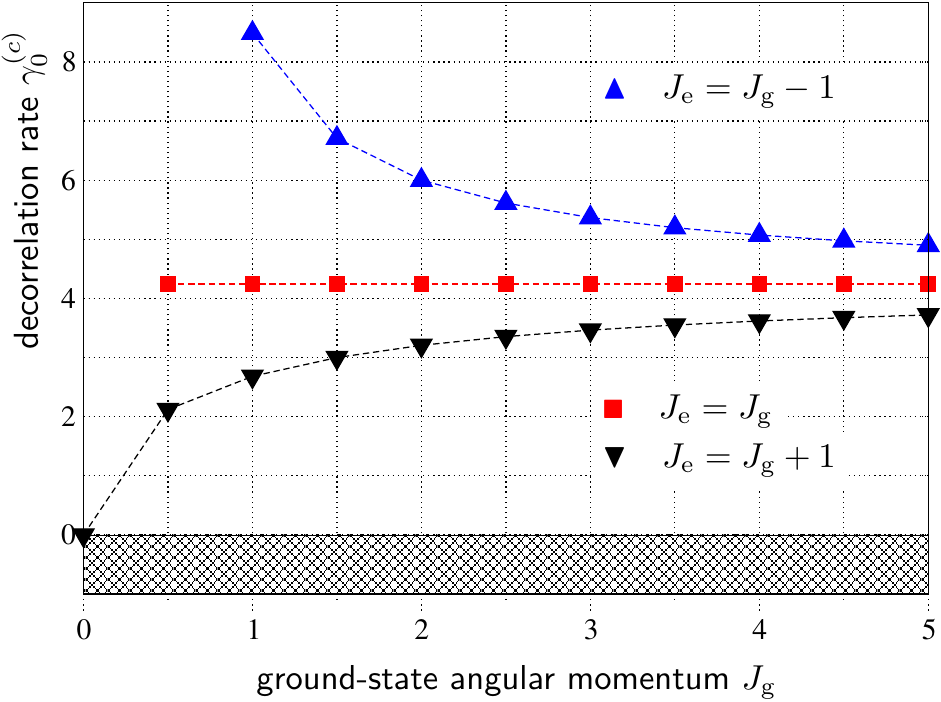}
\caption{(Color online) Decorrelation rate, Eq.~\eqref{gamma0c}, as function of ground-state angular momentum $\Jg$  (in units of $\hbar$) for the 3 possible dipole transition types $|\Je-\Jg|\leq 1$. The hatched area indicates the region $\gamma^{(c)}_0<0$ forbidden by the Rayleigh bound. Except for the non-degenerate case $\Jg=0$ of isotropic dipoles, where the rate vanishes and the Rayleigh law applies, the decorrelation rate is always larger than unity, signalling a rapid loss of correlations. 
}
\label{fig:gammaplot}
\end{center} 
\end{figure} 
Concerning the polarizations, we
assume unpolarized incident light and no polarization selection at
detection, which implies that the weights $t^{(\kappa)}_{ab,ab}=t^{(\kappa)}_{ab}$ given by \eqref{tKab}
are averaged to  
\be
t^{(0)}_{ab} = \frac{2}{3}, \quad t^{(1)}_{ab} =
0, \quad t^{(2)}_{ab} = \frac{1}{3}.
\ee 
The relative fluctuations, Eq.~\eqref{Cabbound}, thus read 
\be
C_{ab,ab} (x) = \left( \frac{2\Gamma_0^\text{(c)}(x) + \Gamma_2^\text{(c)}(x)} {2\Gamma_0(x) + \Gamma_2(x)}\right)^2.
\ee
In the denominator, and for thicknesses $x\gg1$, the diffusive
Goldstone mode $\kappa=0$ with its algebraic decay dominates over the
exponentially decaying mode $\kappa=2$. In the numerator,  
$b_0=1> b_2 =7/10$ implies $\gamma^\text{(c)}_0 <
\gamma^\text{(c)}_2$. Thus, the mode $\kappa=0$ has the slower decrease and
will dominate for thick samples. For large optical thickness $x\gg1$, we therefore find that
the correlations
\be
C_{ab,ab} (x) \sim x^2 \exp\{- 2 \gamma^\text{(c)}_0 x\},  \qquad(\Jg>0),
\label{Cababx}
\ee
are exponentially small, with a rate 
\be
\gamma^\text{(c)}_0 =
\sqrt{3(\Meg^{-1} - 1 )}
\ee 
per mean free path. This decorrelation rate is plotted in Fig.~\ref{fig:gammaplot} as function of $\Jg$ for the 3 different dipole transition types $\Je-\Jg=0,\pm1$. 
For $\Jg=\Je$, one has $\Meg=1/3$, and the prediction $\gamma^{(c)}_0=3\sqrt{2}$ is independent of $\Jg$. For all $\Jg>0$, the rate $\gamma^{(c)}_0$ per mean free path is larger than unity, signalling a rapid loss of correlations. The only case where
correlations should survive is for isotropic dipoles, $\Jg=0,\Je=1$,
where $\gamma^{(c)}_0=0$, and the Rayleigh law \eqref{Rayleigh} applies.

\section{Summary and outlook}
\label{sec:Conc}

We have calculated analytically the short-range speckle correlations measured by light scattering across a dilute cloud of atoms with internal degeneracies, as proposed by \cite{Assaf2007}. These correlations are due to the interference between certain pairs of disjoint path amplitudes. This interference can only occur in the absence of Raman scattering events, which act as path markers. In agreement with our initial claim \cite{Gremaud2008}, we find that these correlations relative to the average intensity are always smaller than unity and cannot exceed the Rayleigh limit, which is only recovered for atoms with a non-degenerate ground state ($\Jg=0$). In transmission across an optically thick slab, the correlations are even predicted to be exponentially small. 
Of course, the exponential reduction \eqref{Cababx}, within the strict protocol proposed in Ref.~\cite{Assaf2007}, does not preclude the enhancement of spectral fluctuations by internal degeneracies, as predicted for an entirely different protocol in Ref.~\cite{Lezama2015}. 

The internal degeneracy of atomic point scatterers is thus found to reduce generic mesoscopic effects such as diffusive intensity correlations and coherent backscattering. As a possible extension of this work, it  would be interesting to analyze whether more long-range correlations---due to paired paths with scatterers in common \cite{AkkermansMontambaux2007}---are affected in a similar way.

\begin{acknowledgments}
The Centre for Quantum Technologies is a Research Centre of Excellence funded by the Ministry of Education and the National Research Foundation of Singapore.
\end{acknowledgments}

\bibliography{intcorrel}

\end{document}